# "New" Challenges for Future C2:
# Commanding Soldier-Machine Partnerships

**Anna Madison, Kaleb, McDowell, Vinicius G. Goecks, and Jeff Hansberger**
U.S. Army DEVCOM Army Research Laboratory
Humans in Complex Systems
Aberdeen Proving Ground, MD
USA
anna.m.madison2.civ@army.mil, kaleb.g.mcdowell.civ@army.mil

**Ceili M. Olney**
Department of Psychology, University of Arizona
Tucson, AZ

Defense Civilian Training Corps Program
Aberdeen Proving Ground, MD
USA

**Claire Ahern**
Virginia Polytechnic Institute and State University
Blacksburg, VA

Department of Defense's Historically Black
Colleges, Universities, and Minority Institute's
Summer Research Internship Program
Aberdeen Proving Ground, MD
USA

**Amar Marathe and Nicholas Waytowich**
U.S. Army DEVCOM Army Research Laboratory
Humans in Complex Systems
Aberdeen Proving Ground, MD
USA

**Christian Kenney**
U.S. Mission Command Battle Lab
Science and Technology Branch
Ft. Leavenworth, KS
USA

**Christopher Kelshaw**
U.S. Mission Command Battle Lab
Futures Branch
Ft. Leavenworth, KS
USA

## ABSTRACT

*Future warfare will occur in more complex, fast-paced, ill-structured, and demanding conditions that will stress current Command and Control (C2) systems. Without modernization, these C2 systems may fail to maintain overmatch against adversaries. We previously proposed robust partnerships between humans and artificial intelligence systems [1], and directly focusing on C2, we introduced how intelligent technologies could provide future overmatch through streamlining the C2 operations process, maintaining unity of effort across formations, and developing collective knowledge systems that adapt to battlefield dynamics across missions [2]. Future C2 systems must seamlessly integrate human and machine intelligence to achieve decision advantage over adversaries while overcoming "new" challenges due to the technological advances driving fundamental changes in effective teaming, unity of effort, and meaningful human control. Here, we describe "new" C2 challenges and discuss pathways to transcend them, such as AI-enabled systems with effective human machine interfaces.*

**"NEW" Challenges for Future C2: Commanding Soldier-Machine Partnerships**

# 1.     Introduction

The transition into a new socio-technological era ushered in by adaptive, intelligent technologies will push future warfare into more complex, fast-paced, ill-structured, and demanding conditions. Along with these changes to the operational environment arise "new" challenges in developing efficient and effective adaptive, intelligent systems that maintain an acceptable level of meaningful human control (MHC), allowing for the ethical and moral accountability required with military use. Two critical assumptions guide this re-imagining of Command and Control (C2) systems. The first is that future operational environments will become more complex, dynamic, and challenging. Increased battlefield lethality will demand distributed C2 systems, and effective command will need the seamless integration of multiple real-time information streams with the ability to conduct operations resilient to disruptions to communication channels. With the rise of the internet of things, personal and connected devices, more data is being produced today than has ever been produced [3]. This trend will continue in future operational environments as both the complexity and number of battlefield assets increase, producing vast amounts of data that can be leveraged and processed for information collection and intelligence processes. The second assumption is that socio-technological progress will fundamentally alter the current human-technology relationship in C2 systems. Future C2 systems enabled by artificial intelligence (AI) technologies will operate at speeds, levels of complexity, and with degrees of intelligence that surpass current human or human-machine integration capabilities. This change in human-machine systems poses challenges for maintaining MHC and unity of effort within and across echelon levels.

As adaptive, intelligent technologies including AI become more integrated into society, we argue that current C2 systems, which stem from the industrial era, will be stressed to the point that without modernization they will fail to adequately perform. We have previously proposed that robust partnerships between humans and AI systems, appropriately applied across factors including task complexity, time availability, and information certainty, potentially can overcome such challenges [1]. Directly focusing on C2, we introduced intelligent technologies in the form of a scalable intelligent course of action (iCOAs) system and provided notional operational examples of how future Soldier-machine partnerships could provide future overmatch through streamlining the C2 operations process, maintaining unity of effort across formations, and developing collective knowledge systems that adapt to battlefield dynamics across missions [2]. Future C2 systems must seamlessly integrate human and machine intelligence to achieve a decision advantage over adversaries. The complexity, dynamic, and ill-defined nature expected in future operational environments necessitates significant human involvement in C2 while acknowledging that technological advances in AI will drive fundamental changes in human roles and actions. In doing so, future AI-enabled C2 systems will enable C2 personnel, including commanders, staff, and liaisons, to increase the number of decisions and achieve a higher decision quality to have more flexible, scalable plans.

Herein, we dive deeper into what it could mean to command using potential Adaptive Intelligent Command and Control (AICOM) capabilities, focusing on meaningful interactions between humans with future adaptive, intelligent technologies. Future adaptive, intelligent technologies that streamline the C2 operations process, for example, capable of automatically and rapidly generating multitudes of potential courses of action (COAs), could fundamentally change how humans and machines partner [4, 5]. We posit that the infusion of human-machine systems into the C2 operations process results in two "new" challenges related to C2 that system developers must address: ensuring effective human-machine teaming and unity of effort.

## 1.1    Purpose

Our objectives here are two-fold. First, we aim to describe the significance of the proposed "new" challenges to future C2, explicitly focusing on MHC. Second, we strive to illustrate potential pathways to address those challenges culminating in describing notional commanders' interfaces and interactions (i.e., describing a system to use by commanders and their staff) for future Soldier-machine C2 systems.

## 2.    Meaningful Human Control (MHC)

The concept of MHC emerged from the increasing concerns over the development and deployment of Autonomous Weapon Systems (AWS) [6]. Historically, weapons were directly controlled by humans, who were fully responsible for their use. However, as technology has advanced, particularly with the advent of AWS, the nature of control has shifted [7]. This has led to fears that these systems might operate without adequate human oversight, potentially compromising legal and ethical standards in warfare [7]. MHC was introduced to address these issues, emphasizing the necessity for human involvement in critical functions of AWS, such as identifying, selecting, and applying force to targets [7].

Two primary premises define MHC: first, that a machine operating and applying force without any human control is broadly unacceptable, and second, that mere human interaction, such as pressing a button based on computer indications without cognitive clarity or awareness, does not constitute acceptable levels of human control [8]. These premises underline the need for human control to be substantial, which is what the term "meaningful" seeks to encapsulate [8]. The framework of MHC includes predictable, reliable, and transparent technology, accurate information for the user, timely human judgment and intervention, and robust accountability mechanisms [6]. However, the inherent imprecision of the term "meaningful" has sparked various interpretations, with different states and organizations proposing divergent standards for what constitutes MHC, ranging from direct human oversight to broader programming-based control mechanisms [9].

Embedding MHC in International Humanitarian Law (IHL) has been discussed as a necessary measure to ensure that human operators remain responsible for AWS actions, maintaining accountability and protecting civilian populations [7]. Integrating MHC throughout the lifecycle of AWS, from design and development to deployment and use, aims to ensure that these systems are predictable, reliable, and transparent, thereby facilitating human understanding and control [6]. Nevertheless, there is a debate over how MHC should be applied in practice, especially in complex or fast-paced scenarios like cyber warfare, where traditional forms of human control may not be feasible [9]. Further complicating this discussion is the lack of consensus on the operationalization of guiding principles at the national level, as identified in international discussions, which emphasizes the need for clear and consistent application of MHC to ensure compliance with applicable laws, notably IHL, and to address ethical concerns [10, 11]. However, without clear definitions and standards, the existing legal framework may become diluted, potentially leading to inadequate protection for civilians and other negative outcomes [8].

Regarding ethical concerns, the International Committee of the Red Cross (ICRC) argues that delegating life-and-death decisions to machines raises profound ethical concerns, particularly regarding the preservation of human dignity and moral responsibility [12]. The ICRC emphasizes that decisions involving the use of force must remain within human agency, as the process of killing or injuring should involve human intent, not be reduced to algorithmic processes. Additionally, the unpredictability and potential unreliability of autonomous systems [13], especially those involving AI and machine learning, present significant risks, further complicating the ethical and legal frameworks within which these systems operate [12].

**"NEW" Challenges for Future C2: Commanding Soldier-Machine Partnerships**

The discussions on MHC are also relevant to future C2 and Soldier-machine partnerships. While the focus has traditionally been on autonomy in critical functions such as target identification and engagement, AI systems embedded in C2 systems and operations processes also present ethical challenges. These systems, which can assist in decision-making by processing vast amounts of information and generating recommendations [4], may inadvertently limit human oversight and moral responsibility, especially as decisions become increasingly complex and distributed across human-machine networks. The risk of reducing human control to mere supervisory roles over AI generated outputs, without adequate understanding or the ability to intervene meaningfully, could raise concerns about accountability in military operations. As such, the principles of MHC should be carefully considered and applied to LAWS and to these broader AI-driven systems encompassing strategic military decisions.

## 3.    Complexity and Today's Operational Environment

U.S. Army C2 personnel, including commanders and staff members, are tasked with managing military operations within highly complex and non-linear operational environments, commonly referred to as open systems [14]. A military unit interacts with its operational environment and the myriad operational variables. These include physical terrain, weather conditions, civilian populations, and the enemy it faces, all of which influence its operations process. The greater the number of variables within the operational environment, the higher the level of complexity the unit faces and the more difficult it is to control mission execution and desired outcomes. Significant battlefield decisions rely on comprehending rapidly changing operational and mission variable behaviors and their complex interactions. As a result, military decision-making, as part of C2, has become too complex and dynamic for individuals to rapidly and accurately understand independently. [15]. Commanders depend heavily on their staff to process information, provide analyses, and make recommendations. Given the volume and complexity of information in modern military operations, this reliance is crucial for timely and effective decision-making. Army C2 processes (i.e., planning, preparing, executing, assessing) have evolved over centuries, integrating lessons learned and technological advancements to support commander decision-making. In their current state of practice, these provide commanders with the necessary tools and frameworks to make acceptably informed decisions under increasing complexity [15].

### 3.1    Characteristics of Meaning Human Control in Today's Army C2

Boutin and Woodcock include a framework of system characteristics for what MHC may entail and what could be considered "appropriate/sufficient" in relation to achieving human control [16]. The characteristics included in this framework are:

- **Understanding** of a system's way of operating and interaction within its environment

- The ability of a human to **evaluate and monitor** the reliability of the systems

- Ability to validate the **usability and serviceability** of the systems

- **Rules of use and rules of engagement** should be explicitly defined and validated

- **Defining and validating precise mission components** (i.e., objectives, targets, restrictions) that the system can use as a framework

- Allowing for the human(s) to exercise judgment vis-a-vis **compliance with appropriate legal regimes** that guide decisions relating to the use of force [16]

Today, several standard C2 practices and procedures manifest these characteristics of MHC included in this framework and allow Army commanders on the modern battlefield to exert acceptable, albeit not absolute, levels of MHC over C2 performance.

### 3.1.1    Understanding

Commanders and staff members execute both confirmation and back briefings to achieve acceptable levels of individual and collective understanding during the planning and execution of a mission. A confirmation briefing ensures that subordinates understand the mission and the commander's intent. During this briefing, the subordinate commanders explain their understanding of the mission, the higher commander's intent, and their plan to accomplish their assigned tasks. The purpose is to confirm that the subordinates' plans are aligned with the higher commander's intent and that any misunderstandings are addressed before execution. Key elements include:

- **Restating Mission and Intent** - Subordinates restate the mission and the commander's intent to ensure they correctly understand them.

- **Critical Tasks and End State** - They outline their tasks and the desired end state.

- **Coordination and Synchronization** - Discuss how their actions will be coordinated and synchronized with other units.

When subordinates restate the mission and the commander's intent, it shows they correctly understood the orders. This mutual understanding builds trust between the commander and subordinates. These briefings promote open communication, where subordinates can clarify doubts and receive feedback. This transparency fosters trust, showing that all parties work towards the same goal with clear expectations. By confirming that their plans align with the higher commander's intent, subordinates demonstrate their commitment to the mission, reinforcing the commander's level of control over their ability to execute the plan effectively. A back briefing is similar to a confirmation briefing but focuses more on the subordinate commanders' detailed explanation of how they plan to accomplish their missions. They emphasize the detailed tactics, techniques, and procedures (TTPs) they will use. The back brief is an opportunity for the higher commander to provide guidance, ensure synchronization, and adjust as necessary. Key elements include:

- **Detailed Plans** - Subordinates present detailed plans, including routes, formations, and actions related to the objective.

- **Risk Management** - They discuss managing and mitigating risks.

- **Resource Requirements** - Identify any additional resources needed.

- **Timelines and Coordination** - Provide timelines for operations and explain how they will coordinate with adjacent and supporting units.

When subordinates explain their plans, it demonstrates that they have thoroughly thought through their actions. This level of detail reassures commanders of their understanding of what is to be achieved and how. Discussing risk management strategies during back briefings demonstrates that subordinates are proactive in identifying and mitigating risks, which builds trust in their judgment and decision-making abilities. By identifying additional resource requirements, subordinates show they are realistic and practical about what is needed for mission success. This honesty and realism foster trust between the commander and subordinates.

Effective coordination and synchronization plans show that subordinates are focused on their tasks and consider the broader operational context. This holistic approach builds trust within the team and with higher command. Both types of briefings are essential for ensuring that all levels of command are synchronized and have a shared understanding of the mission, thereby increasing the likelihood of successful mission execution.

---

**Battle of Nasiriyah (2003)**

During the Battle of Nasiriyah in the Iraq War, the rapid advance and scattered engagements led to communication breakdowns between different Marine units, ground forces, and air support. The reliance on radio communication without direct face-to-face briefings **led to misunderstandings, miscoordination, and an overall lack of understanding.**

The confusion resulted in a friendly fire incident where an A-10 Warthog mistakenly attacked a Marine unit, causing casualties. The incident underscored the importance of clear communication and coordination in joint operations to ensure collective situational understanding.

---

### 3.1.2    Evaluation/Monitoring to Ensure System Reliability & Mission Definition/Validation Frameworks

C2 personnel use feedback mechanisms to adjust their operations. Intelligence reports, reconnaissance, after-action reviews, and communication with higher and lower echelons all provide feedback that informs the evaluation and monitoring of the operation. Execution matrices, decision support tools and templates, and operations orders provide predetermined task specifics, timelines for action, and assignment of responsibilities, along with graphic depictions of required information deemed necessary to allow for effective decision-making and human monitoring of mission execution [17]. Common operating pictures continuously display common data and information shared by multiple commands to assist with shared mission monitoring and evaluation across echelons [17].

---

**Falklands War – Battle of Goose Green (1982)**

During the Falklands War, the Battle of Goose Green saw British forces attempting to capture a key position held by Argentine forces. The initial plan underestimated the strength of the Argentine defenders, and during the battle, the British forces **struggled to assess the progress accurately** due to communication issues and battlefield confusion. Although the British eventually won the battle, the cost was high, including the death of Lt. Col. H. Jones, the battalion commander of the 2nd Battalion, Parachute Regiment. The difficulty in evaluating progress and adapting tactics during the battle highlighted the challenges of mission evaluation in dynamic combat situations.

---

### 3.1.3    Usability/Serviceability of the C2 System

Training and preparation ensure the C2 system—people, networks, processes, and command post operations—is usable and serviceable when employed. The U.S. Army implements several collective training methods to maintain C2 proficiency. For example, a Warfighter exercise is a distributed, simulation-driven, multi-echelon tactical command post-exercise. It places an organization against a live, free-thinking adversary and is designed to train and rehearse units [18].

Staff training exercises allow commanders to train their staff members to perform essential integrating and control functions without deploying the entire unit [18]. For example, a staff exercise (STAFFEX) "trains unit staffs to perform tasks essential to C2 planning, coordination, integration, synchronization, and control functions under simulated operational conditions" [18]. A STAFFEX trains staff to function as an effective team, exchanging information, sharing knowledge, preparing estimates, giving appraisals, making recommendations, preparing orders, issuing orders, and coordinating order execution" [18].

---

**Operation Iraqi Freedom – Post-Invasion Phase (2003-2004)**

After the successful initial invasion, the U.S. and coalition forces struggled with the occupation and stabilization of Iraq. C2 personnel were inadequately prepared for the complexities of post-invasion counterinsurgency missions. The initial **lack of adequate training and preparation of C2 personnel** for post-invasion operations led to widespread instability, a growing insurgency, and a protracted conflict.

---

### 3.1.4    Rules of Use/Engagement & Judgement vis-à-vis International Humanitarian Law

U.S. Army units routinely develop and implement explicit rules of engagement (ROE) that are reviewed and appropriately updated by judge advocates (i.e., military lawyers) for consistency with the Laws of Armed Conflict. U.S. Army C2 personnel are continuously taught basic principles and foundational aspects of the Law of Armed Conflict (LOAC) and rules of engagement that must be adhered to when planning and executing military operations. Further, Army commanders at all levels are responsible for ensuring that all under their command "...operate in accordance with the LOAC and applicable rules of engagement [19].

> **Airstrike on the Doctor's Without Borders Trauma Centre in Kunduz - Afghanistan (2015)**
>
> During the War in Afghanistan, U.S. forces mistakenly bombed a Doctors Without Borders hospital in Kunduz, killing 42 people. The attack on a protected medical facility represented a severe breach of International Humanitarian Law. U.S. forces had mistaken the hospital for a Taliban target due to inadequate intelligence and poor communication. The rules of engagement were not correctly followed, and there was a failure to positively identify the target before striking. The incident led to a significant international condemnation, strained U.S.-Afghan relations, and damaged the credibility of U.S. forces in the region. It also resulted in a pause in the broader military operation and increased scrutiny of ROE compliance, ultimately impacting the overall mission in Afghanistan.

As discussed above, military operations involve many interconnected variables within an open system environment that are beyond a commander's direct and absolute control (e.g., enemy actions, environmental impacts, political and civilian factors, resource constraints). However, several commonly used C2 practices and procedures currently exist that help prepare and allow today's Army C2 personnel to exert acceptable levels of MHC over the conduct of military operations. Incorporating AI into C2 processes in the future will likely change today's practices and procedures for ensuring acceptable levels of MHC, which may result in challenges to MHC and potentially novel methods to maintain it.

## 4. Revolutionizing a Future C2 System: A Series of Evolutions

Historically, large technological disruptions often result from a series of strategic, incremental advances [20]. Apple's iPhone is a quintessential example of such a technology. While revolutionary, the iPhone combined technologies such as touchscreens and miniaturization of components, resulting in a technology much larger than its parts. More recent examples are advancements in AI and machine learning, where innovation occurs through a collective process driven by small technological changes and advancements in shared knowledge across society, leading to major socio-technological breakthroughs [21]. As AI technologies become more ubiquitously used, we imagine that, as a society, *what* we do on the day-to-day may be much of the same, but *how* we execute our tasks may be fundamentally different. In re-envisioning a future C2 system, we likewise expect this to be true. A future C2 system enabled by AICOM capabilities will still perform all aspects of today's C2 system (potentially more), but how it accomplishes C2 processes, procedures, and functions will significantly be changed by technologies.

The advancement of a future C2 system enabled by AICOM capabilities will most likely occur through a series of evolutionary steps that, over time, perhaps decades, as a whole, will have resulted in a larger revolution of the (current 2024) C2 system (see Figure 1). The process of integrating technology itself generally takes time, while the integration of technology specifically into military contexts is usually a lengthy process due to the relatively higher standards for reliability, security, and effectiveness, along with the need for extensive testing, doctrinal changes, training, and alignment with strategic goals [22, 23]. As AICOM capabilities are developed, evolved, tested, and evaluated, they will be integrated into C2 functional and integrating cells performing the C2 operations and integrating processes by replacing or augmenting current processes, procedures, and functions but not the C2 personnel therein who are critical for maintaining MHC.

**"NEW" Challenges for Future C2: Commanding Soldier-Machine Partnerships**

We expect, as the system evolves, for a suite of specific AICOM capabilities to emerge designed to augment each Army warfighting functional cell (i.e., intelligence, fires, sustainment, protection, mission command, and movement and maneuver) and integrating cell (i.e., current operations, future operations, and plans) all performing C2-related tasks that allow for higher quality decisions made by commanders leading to decision advantages over adversaries. While some aspects of the C2 operations process may be automated completely to operate with little human oversight (e.g., preparing orders and plans), other elements (e.g., critical decisions, assessments requiring judgment) will be retained by human C2 personnel (i.e., commanders, staff members, liaisons) or other collaborators to ensure adequate levels of MHC. The resulting C2 system enabled by AICOM capabilities should result in a streamlining and subsequent compression of the C2 operations process, producing more actionable data analytics, decision recommendations, and C2 products in a given period than could ever be achieved by humans alone. We posit that a consequence of this AICOM-enabled C2 system is a reduction in how many C2 personnel are required for overmatch in an operational environment.

## 5.    Effective Human-Machine Partnerships

Establishing human-machine partnerships with adaptive, intelligent technologies will present many teaming considerations for the future conduct of C2. While the nature of these human-machine partnerships may be similar to human partnerships [24], managing these new types of partnerships introduces teaming challenges [25]. The reduction in the overall number of human C2 personnel by enabling one or several humans to efficiently and more effectively complete the C2 operations process allows for more traceability in terms of MHC, but a by-product of the integration of AI and adaptive, intelligent technologies is the drastic increase in complexity of human-machine partnering.

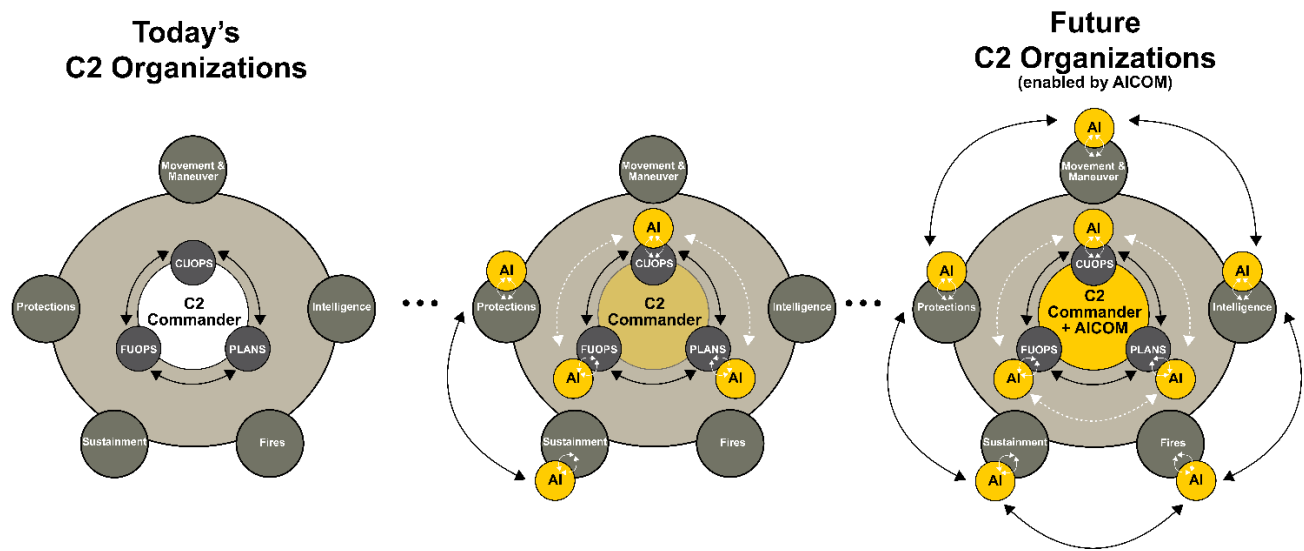

**Figure 1: The transformation of the Command and Control (C2) system through adaptive, intelligent technologies will start with incremental changes within existing functional and integrating cells, as depicted on the left. As AICOM capabilities are developed and tested, they will be gradually integrated into this organizational structure. Over time, as adaptive and intelligent technologies evolve, we anticipate they will become pervasive across these organizations, becoming an integral part of the overall C2 system.**

## "NEW" Challenges for Future C2: Commanding Soldier-Machine Partnerships

While human-machine partnerships can encompass a landscape of tasks, the nature of the human-AI interaction depends on task and task complexity, decision timescales, and information certainty [1]. However, only considering the capabilities of AI (e.g., sensing, reasoning, analysing) or humans (e.g., handling levels of uncertainty or ambiguity) limits the perceived roles of each. Instead of placing humans and AI in their own categories and determining their respective roles based on their perceived capabilities, the nature and expectations of a human-AI partnership should be task-dependent. In complex data-processing tasks, for example, that require quick analysis of a vast amount of data, an AI system may be able to process the data efficiently. Still, human intelligence may be needed to provide contextual interpretation. As the composition of C2 personnel and organizations shifts with the incorporation of AICOM systems to enhance human-machine partnerships through more optimal selection [5], critical aspects of teaming, such as communication, creating and maintaining situational awareness (SA), validating recommendations, and developing and maintaining unity of effort, must be adapted to include emerging intelligent systems.

## 5.1    Team Management Challenges

Current C2 organizations (see the left side of Figure 1) are composed of C2 personnel augmented by "computer" technology (rather than "intelligent" technology), leveraging mostly human intelligence to oversee, aggregate, and analyze data and intelligence so that commanders can make timely and quality decision (but see [26]). Driven by human teams organized by Army war-fighting functional and integrating cells, information is exchanged within a given functional cell and integrated across the cells to create shared mental models[1] and shared SA[2], which have been shown to influence team effectiveness and performance [31, 30]. In human-agent interaction, an agent's transparency is essential for maintaining or enhancing the user's SA [32, 33] These human teams' shared mental models and SA are facilitated mainly by a common language used for information exchange (e.g., conversations, chat, commander's briefings, and visualizations), where quantity does not always predict quality [34].

In human teams and human-machine partnerships, effective collaboration heavily depends on developing shared mental models [27] and shared SA [35]. However, how human and human-AI teams achieve acceptable levels of shared mental models, and SA are fundamentally different due to differences in how humans and AI communicate and process information. Humans rely on cognitive processes like attention, working memory, and abstract thinking. If a question arises, one can ask the human to explain or show the analysis leading to a decision or final outcome. AI, however, operates on algorithms that can be "black boxes" [36] and may not be transparent or easily understood by humans, leading to miscalculating trust[37]. A misalignment in trust can lead to reduced SA, mainly when humans are tasked with monitoring AI systems, and cause them to become unintentionally "out-of-the-loop" and unable to intervene effectively [38, 35].

Compounding these "normal" teaming challenges, AI's ability to integrate, process, and analyze data rapidly to enhance the speed and quality of decision-making, crucial in fast-paced operational environments, comes with the complexity of ensuring that AICOM systems are transparent, human teammates understand decisions, and acceptable MHC can be maintained. At the core of this dynamic is creating and maintaining calibrated and sufficient levels of trust in ways that minimize distrust. Too little trust could lead to errors due to disuse, while blind trust may lead to human miscalculations with assessing risk and a general overreliance on the machine [39, 40]. Calibrated trust solely among humans evolves through shared experiences and situational understanding and can be developed through training [41, 42]. In contrast, trust in human-AI teams is more fragile and adds to the reliance on the transparency and reliability of the AI system. Human-AI teams must, therefore, prioritize the development of AI systems that are accurate and explainable so that human teammates can build trust in the AI and AICOM capabilities. While trust enables the appropriate action and MHC over an AI or automated system, bias can still be a prevalent problem, along with how best to resolve moral and ethical decisions [43, 44] in both human and human-AI partnerships. Moreover, the dynamic nature of AI and future adaptive, intelligent technologies—capable of altering their behavior through learning—adds complexity to maintaining accurate shared mental models and consistent situational awareness (SA). The key challenge will be developing human-AI teams that harness the adaptability of both humans and AI, ensuring that the team can respond effectively to evolving conditions while maintaining a unified effort.

---

[1] Shared mental models can be defined as individual team members' mental models allowing for similar understanding each other's roles, capabilities, and intentions [27].

[2] Situation Awareness (SA) as a mental model of the current environment developed through the perception of environmental elements (level 1), the comprehension of their meaning (level 2), and the projection into the near future (level 3) [28, 29]. Shared situation awareness (SA) can be thought of as a higher level of individual SA and coordination among team members [30] leading to a common understanding of the current known situation [31].

## 5.2 Maintaining Unity of Effort

Unity of effort is a critical principle in military operations, ensuring that all mission elements are coordinated and directed towards a common objective [45, 46]. Three contributing teaming components to achieving unity of effort are communication and coordination, shared mental models and SA, and common future objectives and planning goals. Currently, the Army's principle of war on unity of command puts a commander responsible as the essential component for success in an operation [45, 46]. With a central point of control, this is, by and large, a final stop-gap to ensure unity of effort is achieved and to maintain MHC. However, before the commander, an organizational structure allows Warfighters to communicate and coordinate in deliberate and procedural manners, facilitating the development of shared mental models and SA. Through common doctrine, training, tactics, techniques, and procedures, shared mental models can be developed and shaped to better predict other Soldier's behaviors and/or how to operate in response to maintain a unity of effort and achieve mission success.

The introduction of intelligent technologies into the C2 operations process creates two unity of effort challenges as a result of the overwhelming complexity and rapid pace of the operational environment. First, creating a shared, distributed SA across multi-echelons in response to the rapidly changing operational environment will be difficult. Inundated with information and data analytic capabilities, the compression of timescales and information complexity will make understanding the battlefield both easier but more dense. Second, with the rise of AI and AICOM technologies, we expect an increase in communication disruptions, making maintaining a common operation picture or shared SA across distributed groups or units a necessary capability. While the intelligent technologies themselves hold the key to overcoming these changes, the system is only as good as it is usable. Next, we discuss approaches and technologies that can be integrated and employed to overcome these "new" C2 challenges related to effective human-machine teaming and unity of effort.

## 5.3 Approaches and Technologies to Ensure Effective Human-Machine Teaming and Unity of Effort

While sound systems and user designs will not fix all the challenges in a future C2 system enabled by intelligent systems, they will be a component in solving many of the above-described challenges. By enabling effective teaming, unity of effort is far easier to achieve and maintain. An emerging approach to facilitate efficient workflows between humans and AI is known as team design patterns. Team design patterns are an efficient way to manage workflows in human-machine partnerships, providing generic reusable behaviors across the team that can be combined based on the task to accomplish a goal and a way to integrate humans and machine intelligence seamlessly [47, 48]. Within a future C2 system, team design patterns could be used to modernize processes and procedures of functional and integrating cell members within a headquarters staff. Developing decision-making frameworks that integrate AI-driven insights with human oversight can help balance the strengths of both [1].

Communication is essential to coordinated teamwork, shared mental models, and SA. Improving communication and information exchange between human-machine partners can facilitate the impact of system transparency [35] and explainability [49] by aligning to tasks, required human oversight, and system reliability [32, 33]. This, in turn, helps to build trust within human-AI partnerships [49]**by** providing information in a way that helps shape more accurate mental models of the AI, allowing the human teammates to understand and rely on AI decisions [27]. Transparency in AI systems, mainly through real-time and future-oriented displays or interfaces, is crucial for supporting higher levels of SA [35]. But without proper system and interface design that is intuitive and easy to interpret [50], these approaches to increase communication, shared mental models, and SA fall flat [37]. In doing so, future intelligent-enabled C2 system design approaches may transcend current limitations to MHC, which often dichotomize control between a system (i.e. autonomy vs. human operator) [51, 52] by better representing the overall human-AI or -machine system status more accurately allowing for application to military processes and contexts that are complex, dynamic, and multidimensional.

Finally, where technology and system design fail, the Army (and other large organizations) are experts in training personnel to operate with the organizational structure and the provided tools and technologies to accomplish mission objectives. Recent work suggests that human users' adaptability and flexibility help them overcome imperfect AI's limitations, such as in code translation tasks [53]. Importantly, evidence suggests components important in human-machine partnerships can be developed or trained, for example, reasoning over quantified uncertainty [54] or influencing trust in human-machine teams since it can evolve over time [55]. Finally, enhancing adaptability in human-AI teams may involve training human team members to better understand and work with AI systems and designing AI systems that are more attuned to human needs and behaviors through user-centered design approaches. In the remainder of the paper, we combine these approaches and technologies into a Future Battlefield Visualization to illustrate a pathway to address "new" challenges related to human-AI partnerships.

# 6.     Future Battlefield Visualization: Enhancing C2 through User Interface Design

Fundamental design principles for ensuring effective future human-AI partnerships and unity of effort include organizing key information around goals, supporting higher-level SA processes (e.g., comprehension and projection), and minimizing cognitive workload by presenting information in a coherent and easily digestible manner [35]. User interfaces for future C2 personnel interfaces should support the dynamic updating of shared mental models through transparency by clearly communicating changes in the AI's capabilities, behavior, or potential biases. Real-time and future displays should be prioritized to ensure humans remain in control and can intervene effectively when needed while ensuring that information is salient and integrated to prevent it from getting lost in busy displays. Properly designed interfaces containing the right information with the correct user experience can lead to efficiencies in decision-making, leading to increased transparency, which influences trust and shared mental models. Following a similar process as the Situation Awareness Oriented Design process [38][35], which focuses on developing interfaces to support fundamental design principles, OpsVision was designed as an interface for future C2 personnel, staff, and knowledge workers.

OpsVision, as illustrated in Figure 2, leverages spatial computing to create an intuitive, immersive, and contextually aware interface for enhancing information comprehension and decisionmaking capabilities. Spatial computing represents a paradigm shift in how we interact with digital information by merging the physical and digital worlds. At its core, spatial computing encompasses technologies that enable the seamless interaction of virtual and real environments, often through a combination of augmented reality, virtual reality, and mixed reality interfaces. In the face of large amounts of data, information overload, and operational environment complexity, relevant C2 information must be conveyed to maximize C2 effectiveness and system understanding to exert MHC fully. OpsVision augments unity of effort by enhancing shared SA, supporting the commander's visualization and planning of future objectives, and improving communication and coordination. By integrating real-time data into a cohesive and intuitive spatial display, OpsVision provides a comprehensive view of the operational environment, allowing commanders to make informed decisions and respond quickly to changing conditions. The structured framework for strategic planning and goal setting enables detailed exploration and simulation of different scenarios, ensuring that all efforts are aligned toward achieving mission objectives. The shared visual platform and integrated communication tools facilitate effective collaboration and coordination, maintaining unity of effort across all mission elements. As we delve further into this section, we explore how these capabilities can be harnessed to create a future battlefield environment that is more responsive, adaptive, and effective in meeting the complex challenges of modern warfare.

# WORK MODE          # INTEGRATION MODE

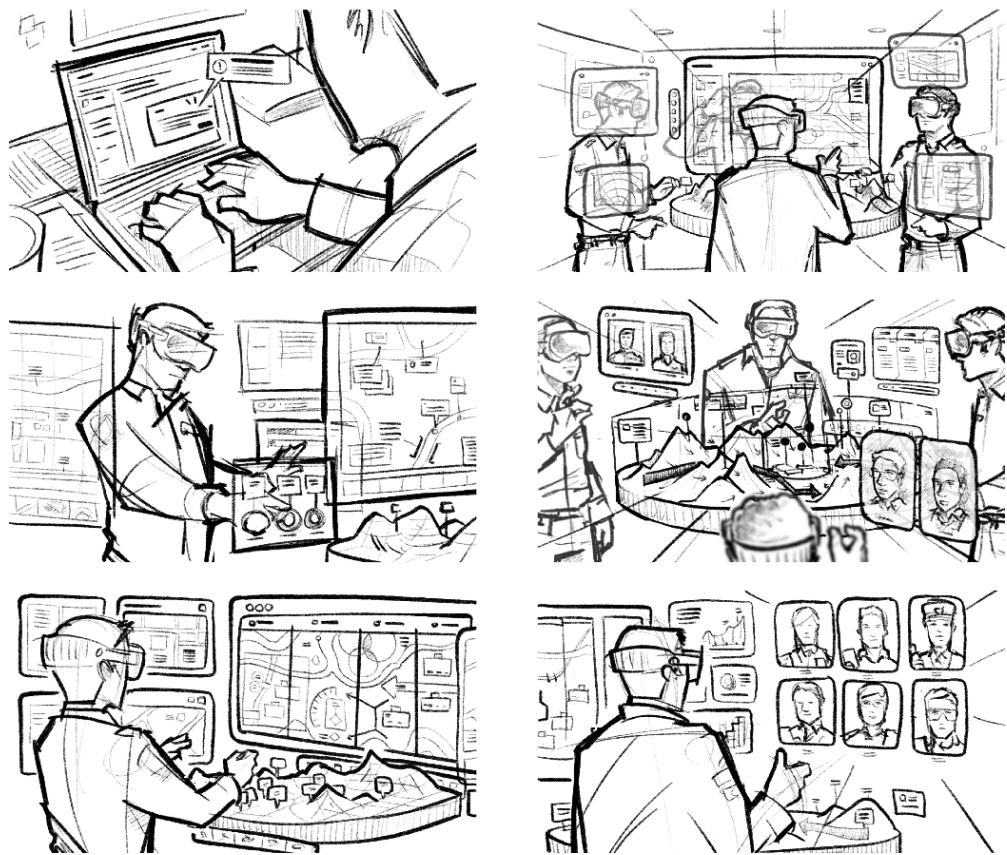

**Figure 2: OpsVision. A short overview of a commander and C2 personnel interacting with a future C2 personnel's interface with optimized user interactions leveraging multiple input modalities (e.g., text, voice, gesture) and interface design that facilitates intuitive, efficient information comprehension within a spatial computing environment. Different modes are shown for work (left column) and integration (right column) user workflows.**

## 6.1    Optimizing Information for Enhanced C2 Performance

C2 information and data are inherently complex, encompassing many data types, sources, and contexts. This complexity manifests in both breadth, covering a wide range of operational aspects from logistics to intelligence, and depth, involving detailed and nuanced data that requires careful interpretation and analysis. As new technologies, such as AI, continue to evolve and integrate into C2 systems, this system and information complexity are expected to scale exponentially. AI can accelerate data processing and provide real-time insights. Still, it also reduces the time available for decision-making, necessitating a more efficient way to manage and interpret vast amounts of information quickly and accurately.

C2 systems can improve SA and decision-making processes by presenting data in a more accessible and interpretable format. One effective method for managing information complexity is through narrative visualization, a technique that structures information in a coherent, story-like format, making it easier to understand and follow [56, 57]. This approach leverages the human capacity for processing stories, which are naturally engaging and memorable, to enhance comprehension and retention of complex information [57]. Narrative visualization integrates data with narrative elements to create a cohesive and compelling presentation by organizing data into a structured sequence to guide the information logically and intuitively.

This narrative approach, coupled with efficient structures for organizing information, such as the martini glass structure[3], offers several potential benefits. The martini glass structure enhances comprehension by structuring information in a narrative format, making complex data more understandable and accessible. This improved clarity supports more informed decision-making, as the logical flow of information ensures that all relevant data is considered, reducing the cognitive load on decision-makers and allowing for quicker, more accurate decisions. The martini glass structure is inherently flexible and adaptable, accommodating different types of C2 data and scenarios. This versatility is crucial in dynamic environments where information needs to be rapidly interpreted and acted upon. When integrated into a spatial interface like OpsVision, the martini glass structure can leverage the strengths of spatial computing to present information in a visually intuitive and interactive manner. This integration further enhances the user experience and effectiveness of the C2 system by providing an engaging and interactive way to explore and understand data.

In the context of tasks, processes, and procedures related to the C2 operations process, two primary modes of interaction are critical for effective mission planning and execution: work and integration mode. OpsVision incorporates these different user modes to reflect the distinct phases of individual and collaborative tasks performed by C2 personnel, seamlessly transitioning between individual work and collaborative sessions to enhance overall workflow efficiency and coordination.

- **Work Mode**: This mode is characterized by focusing on specific individual C2 personnel duties and responsibilities. Each member utilizes available resources, including data analysis tools, communication devices, and planning software, to conduct detailed work relevant to their role. This phase involves intensive data processing, planning, and coordination efforts requiring concentration and specialized information access. Much of the C2 personnel's time is spent in this mode, ensuring thorough preparation and detailed task execution.

- **Integration Mode**: This mode is characterized by its collaborative features that enable periodic communication across co-located distributed C2 personnel or come together to integrate their individual efforts, typically around a shared resource such as a map or a digital display. These sessions are essential for aligning strategies, updating SA, and creating unity of effort. This collaborative effort is often dramatized in media, where team members gather around a physical map or a central display to discuss plans and strategies.

## 6.2    Designing for Spatial Interfaces

OpsVision employs the principles of peripersonal and extrapersonal spaces to optimize the user interface for both work and integration modes to enhance usability and interface effectiveness. Spatial interfaces naturally map to the concepts of peripersonal[4] and extrapersonal[5] spaces, which are fundamental to human spatial cognition and interaction. Research in neuroscience and human factors has shown that these spatial categories are deeply embedded in how we perceive and interact with our environment. Peripersonal space is associated with direct manipulation and immediate interactions, making it ideal for tasks that require hands-on control and quick access, such as real-time operational monitoring [58]. Extrapersonal space, conversely, is suited for observation and planning, as it allows for the integration of broader contextual information and strategic considerations [59].

---

[3] In the martini glass structure, information is presented in three phases: (1) a broad introductory phase providing a situation overview; (2) an exploratory phase that allows for a deeper dive into detailed data and analysis and engagement with information; and (3) a conclusion phase that synthesizes the explored data into a coherent summary and actionable insights providing recommendations and plans to ensure comprehensive understanding.

By effectively integrating peripersonal and extrapersonal spaces, OpsVision enhances the C2 spatial interface's ability to support individual and collaborative tasks. This approach aligns with human spatial cognition and improves the efficiency and effectiveness of military C2 operations. Understanding the spatial arrangement of information is essential for optimizing workflow and interaction, highlighting the importance of the right content and data streams to fully leverage OpsVision's capabilities for optimal C2 operations. In OpsVision, peripersonal space includes all the information and control elements within arm's reach, arranged in a console structure that supports both standing and sitting positions. This setup is adaptable in stationary environments, such as command post nodes and mobile environments. The console structure allows for direct interaction with critical data feeds, control panels, and communication tools, facilitating real-time operational monitoring and decision-making. By positioning essential information within easy reach, OpsVision ensures commanders can quickly access and manipulate data, enhancing their ability to respond to dynamic situations. The extrapersonal space in OpsVision mimics traditional white and cork boards found in command post nodes but with significant enhancements. This space displays running estimates and other reference materials for different C2 areas, providing persistent and real-time updates. Unlike static whiteboards that require manual updates, OpsVision's extrapersonal displays automatically reflect the latest information, ensuring all staff members can access the most current data. This dynamic and interactive space supports strategic planning by presenting a comprehensive view of the operational environment, enabling commanders to visualize large-scale maps, resource allocations, and long-term plans.

## 6.3 Interface Components

Spatial interfaces provide immersive mixed-reality environments where information can be presented in more interactive and engaging ways through visualizations coupled with multi-modal interactions. The primary attributes of spatial interfaces are their enhanced visualization capabilities beyond traditional 2D displays, such as computer monitors and mobile screens, which limit the depth and context that can be conveyed. In contrast, spatial interfaces can present data in three dimensions, providing a more comprehensive view of complex information. For military commanders, this means the ability to visualize battlefield scenarios in real-time, with a depth of detail previously unattainable. This three-dimensional (3D) visualization enables a more intuitive understanding of terrain, troop movements, and logistical considerations, fostering quicker and more informed decision-making. Additionally, spatial visualization can integrate various data streams, such as intelligence reports, satellite imagery, and real-time sensor data, into a cohesive view, further enhancing SA. Spatial interfaces also offer advanced multi-modal interaction capabilities, significantly improving user experience and operational efficiency. Multi-modal interactions refer to using multiple methods for interacting with digital information, such as voice commands, gesture recognition, eye tracking, and traditional inputs like keyboards and touchscreens. For military applications, commanders and soldiers can interact with spatial computing systems in the most intuitive and efficient manner possible, depending on the context and environment. For instance, gesture recognition can provide an effective alternative in a noisy battlefield environment where voice commands may be impractical. Switching seamlessly between different modes of interaction ensures that users can always maintain control and access crucial information, regardless of the operational conditions.

---

[4] Peripersonal space refers to the area immediately surrounding the body, where objects can be reached and manipulated directly.

[5] Extrapersonal space encompasses the area beyond immediate reach, where objects must be observed and interacted with at a distance.

## 6.4    C2 Personnel Interfaces

Effective communication and coordination are vital for achieving and maintaining unity of effort, ensuring that all mission elements are synchronized and working towards the same goals. OpsVision augments communication and coordination by providing a shared visual platform that facilitates collaboration among team members. The spatial interface supports real-time updates and interaction, allowing staff to communicate and coordinate more effectively. For instance, commanders and staff can gather around a central display during integration mode to discuss plans, share updates, and make coordinated decisions. The ability to visualize information collectively enhances mutual understanding and fosters a unified approach to operations. OpsVision's integration of various communication tools, such as voice and gesture controls, further streamlines the coordination process, enabling seamless information exchange, collaboration, and accurate shared SA within and across echelons.

SA is the foundation of effective decision-making in military operations, requiring a comprehensive understanding of the present operational environment, including the location and status of friendly and adversarial forces, terrain features, and other relevant factors and the future projection of how to act in response. OpsVision enhances SA by integrating real-time data from multiple sources into a cohesive spatial display, where Commanders can visualize the battlefield in 3D, providing a more intuitive and detailed understanding of the operational landscape. The ability to overlay different data streams, such as satellite imagery, drone feeds, and sensor inputs, ensures that all relevant information is readily accessible and continuously updated. This enriched SA allows commanders to make more informed decisions and respond swiftly to dynamic battlefield conditions to promptly begin planning.

Future objectives and planning goals guide military operations toward successful outcomes. These goals require careful planning and strategic foresight, considering both immediate and longterm factors. OpsVision supports this component of UoE by providing a structured framework for strategic planning and goal setting. The spatial interface allows for the visualization of future objectives and the exploration of various COAs within a 3D context. Commanders can simulate different scenarios, assess potential outcomes, and adjust plans accordingly. This capability is particularly valuable for COA development and MDMP, where understanding different strategies' spatial and temporal implications is crucial. By enabling detailed planning and visualization of future goals, OpsVision enhances the alignment of efforts toward achieving mission objectives.

## 6. 5    Interface Data Requirements: Core Information Types for C2 Operations

To maximize the effectiveness of future C2 systems, it is essential to identify and incorporate the primary types of data that are critical to planning and execution efforts. By seamlessly integrating geospatial, temporal, and contextual information, OpsVision provides a holistic view of the operational environment, enabling commanders and staff to achieve superior SA and decision-making capabilities. This comprehensive approach to data integration supports the dynamic and complex nature of modern military operations, ensuring that all relevant information is readily accessible and actionable within the C2 spatial interface.

- **Geospatial information** is foundational to C2 operations, providing the spatial context necessary for understanding the physical environment in which missions are conducted. This includes detailed maps, terrain analysis, and the positioning of both friendly and adversarial forces. High-resolution satellite imagery, Geographic Information System data, and real-time location tracking of assets are integral components of geospatial information. Visualizing and manipulating this data within a spatial interface like OpsVision allows commanders to assess terrain advantages, identify potential obstacles, and plan routes effectively. For example, during a mission, geospatial data can highlight safe pathways, evacuation routes, and optimal positioning for troops and equipment, ensuring that strategies are grounded in an accurate understanding of the environment [60].

- **Temporal information** is another critical data type, encompassing all time-related aspects of mission planning and execution. This includes timelines, schedules, and synchronization of activities across different units and operational phases. Effective temporal data management ensures that all actions are coordinated and timed precisely to achieve strategic objectives. In OpsVision, temporal information can be integrated with geospatial data to provide a comprehensive view of how events unfold over time. Commanders can use this capability to plan complex operations that require precise timing, such as coordinated strikes, logistical resupply missions, and synchronized maneuvers. Real-time updates to temporal data also allow for dynamic adjustments in response to changing battlefield conditions, enhancing the agility and responsiveness of C2 operations.

- **Contextual information** encompasses a wide range of data that provides additional layers of meaning and relevance to geospatial and temporal information. This includes intelligence reports, environmental conditions, cultural and socio-political factors, and mission-specific details such as objectives and rules of engagement. Contextual information is crucial for understanding the broader implications of actions and decisions within the operational environment. In OpsVision, contextual data can be overlaid on geospatial and temporal displays, enriching the information landscape and enabling commanders to make more informed decisions. For instance, integrating weather forecasts with terrain data can help plan troop movements, while socio-political intelligence can inform engagement strategies and mitigate risks associated with local populations [61].

In summary, OpsVision demonstrates the transformative potential of spatial computing in C2 systems by integrating effective design elements that aim to enhance SA, support strategic planning, and improve communication and coordination. By leveraging advanced visualization techniques, multi-modal interactions, and structured information frameworks, such as narrative visualizations and the martini glass structure, OpsVision optimizes workflow and decision-making processes. This comprehensive approach aligns with human cognitive strengths. It maximizes the capabilities of emerging technologies, ultimately fostering greater coherence, efficiency, and effectiveness in military operations and contributing to the success of mission objectives.

# 7. Conclusions

In conclusion, as we transition into a new socio-technological era characterized by integrating adaptive, intelligent technologies, the landscape of future warfare will be fundamentally altered. The complexity, speed, and demands of modern battlefields will require C2 systems that are both efficient and capable of maintaining a meaningful level of human control. The ethical and moral accountability embedded in military operations demands that these advanced systems do not operate in isolation but work in tandem with human commanders and staff. While incorporating AI into C2 processes promises to revolutionize how decisions are made on the battlefield, it also introduces significant challenges. These include ensuring that AI systems are accurate, transparent, and explainable and fostering calibrated trust within human-machine teams. The future of C2 will likely involve a gradual evolution of practices and procedures driven by advancements in AI and AICOM capabilities. As these technologies develop, they will enhance SA, decision-making, and operational effectiveness, potentially leading to a decision advantage over adversaries. However, the successful integration of AI into C2 systems hinges on overcoming barriers related to trust, transparency, and the explainability of these systems. Establishing effective human-machine partnerships will be critical, requiring careful consideration of how trust is calibrated and how shared mental models are developed. Future interface designs must prioritize these factors to ensure that human-AI teams can operate with unity of effort and shared SA. Ultimately, the evolution of C2 systems enabled by AI and AICOM technologies offers the potential for a new level of intelligence and decision-making capability. Yet, the ethical implications and the necessity for maintaining MHC underscore the importance of continuing to develop these systems with a focus on human-centric values, ensuring that they serve as practical tools in the hands of military personnel rather than autonomous entities beyond human control. The path forward will require a careful balance of innovation, ethical considerations, and preserving human judgment in military operations.

# 8. Acknowledgments

# 9. References

[1] J. S. Metcalfe, B. S. Perelman, D. L. Boothe, and K. Mcdowell, "Systemic oversimplification limits the potential for human-AI partnership," *IEEE Access*, vol. 9, pp. 70242–70260, 2021.

[2] K. McDowell, E. Novoseller, A. Madison, V. Goecks, and C. Kelshaw, "Re-envisioning command and control," in *International Conference on Military Communication and Information Systems (ICMCIS)*, 2024.

[3] Statista, "Volume of data/information created, captured, copied, and consumed worldwide from 2010 to 2025," 2023. Accessed: 2024-08-11.

[4]  V. G. Goecks and N. Waytowich, "COA-GPT: Generative pre-trained transformers for accelerated course of action development in military operations," in *International Conference on Military Communication and Information Systems (ICMCIS)*, 2024.

[5]  A. Madison, E. Novoseller, V. G. Goecks, B. T. Files, N. Waytowich, A. Yu, V. J. Lawhern, S. Thurman, C. Kelshaw, and K. McDowell, "Scalable interactive machine learning for future command and control," in *International Conference on Military Communication and Information Systems (ICMCIS)*, 2024.

[6]  H. M. Roff and R. Moyes, "Meaningful human control, artificial intelligence and autonomous weapons," in *Briefing Paper Prepared for the Informal Meeting of Experts on Lethal AuTonomous Weapons Systems, UN Convention on Certain Conventional Weapons*, 2016.

[7]  T. Chengeta, "Defining the emerging notion of" meaningful human control" in weapon systems," *New York University Journal of International Law and Politics*, vol. 49, no. 3, pp. 833– 890, 2017.

[8]  R. Moyes and H. Roff, "Key elements of meaningful human control," in *Background Paper Prepared for the Informal Meeting of Experts on Lethal Au-Tonomous Weapons Systems, UN Convention on Certain Conventional Weapons*, Article 36, 2016.

[9]  R. Crootof, "A meaningful floor for" meaningful human control"," *Temple International & Comparative Law Journal*, vol. 30, no. 1, p. 53, 2016.

[10] United Nations Office at Geneva, Group of Governmental Experts on Emerging Technologies in the Area of Lethal Autonomous Weapons Systems, "Report of the 2018 session of the group of governmental experts on emerging technologies in the area of lethal autonomous weapons systems." United Nations Document CCW/GGE.1/2018/3, October 2018. Available at: https://documents.un.org/access.nsf/get?OpenAgent&DS=CCW.1/2018/3&Lang=E.

[11] United Nations Office at Geneva, Group of Governmental Experts on Emerging Technologies in the Area of Lethal Autonomous Weapons Systems, "Commonalities in national commentaries on guiding principles." Working Paper, CCW/GGE.1/2020/WP.1, October 2020. Available at: https://documents.un.org/access.nsf/get?OpenAgent&DS=CCW/GGE.1/2020/WP. 1&Lang=E.

[12] International Committee of the Red Cross, "Ethics and autonomous weapon systems: An ethical basis for human control?," 2018.

[13] National Highway Traffic Safety Administration, "Standing General Order 2021-01, Second Amended - NHTSA," 2021. Available at: https://www.nhtsa.gov/sites/nhtsa.gov/ files/2023-04/Second-Amended-SGO-2021-01_2023-04-05_2.pdf.

[14] J. Gharajedaghi, *Systems thinking: Managing chaos and complexity: A platform for designing business architecture*. Elsevier, 2011.

[15] D. R. Worley, "Understanding commanders' information needs," tech. rep., RAND, 1989.

[16] B. Boutin and T. Woodcock, "Aspects of realizing (meaningful) human control: a legal perspective," in *Research Handbook on Warfare and Artificial Intelligence*, pp. 179–196, Edward Elgar Publishing, 2024.

[17] Headquarters, Department of the Army, *Army Field Manual 5-0: Planning and Orders Production*, 2022.

[18] Headquarters, Department of the Army, *Training*, 2021.

[19] Unted States Army, United States Marine Corps, The Judge Advocate General's Legal Center and School, *The Commander's Handbook on the Law of Land Warfare*, 2019.

[20] W. B. Arthur, *The nature of technology: What it is and how it evolves*. Penguin UK, 2010.

[21] M. Muthukrishna and J. Henrich, "Innovation in the collective brain," *Philosophical Transactions of the Royal Society B: Biological Sciences*, vol. 371, no. 1690, p. 20150192, 2016.

[22] W. R. Murray and A. R. Millett, *Military innovation in the interwar period*. Cambridge University Press, 1998.

[23] S. P. Rosen, *Winning the next war: Innovation and the modern military*. Cornell University Press, 1991.

[24] J. B. Lyons, K. Sycara, M. Lewis, and A. Capiola, "Human–autonomy teaming: Definitions, debates, and directions," *Frontiers in Psychology*, vol. 12, p. 589585, 2021.

[25] F. L. Oswald, M. R. Endsley, J. Chen, E. K. Chiou, M. H. Draper, N. J. McNeese, and E. M. Roth, "The national academies board on human-systems integration (bohsi) panel: Human-ai teaming: Research frontiers," in *Proceedings of the Human Factors and Ergonomics Society Annual Meeting*, vol. 66, pp. 130–134, SAGE Publications Sage CA: Los Angeles, CA, 2022.

[26] S. Mohsin, "Inside project maven, the us military's ai project," 2024.

[27] R. W. Andrews, J. M. Lilly, D. Srivastava, and K. M. Feigh, "The role of shared mental models in human-ai teams: a theoretical review," *Theoretical Issues in Ergonomics Science*, vol. 24, no. 2, pp. 129–175, 2023.

[28] M. R. Endsley, "Situation awareness," *Handbook of Human Factors and Ergonomics*, pp. 553–568, 2012.

[29] M. R. Endsley, "Situation awareness," *Handbook of Human Factors and Ergonomics*, pp. 434–455, 2021.

[30] J. C. Gorman, N. J. Cooke, and J. L. Winner, "Measuring team situation awareness in decentralized command and control environments," in *Situational Awareness*, pp. 183–196, Routledge, 2017.

[31] L. D. Saner, C. A. Bolstad, C. Gonzalez, and H. M. Cuevas, "Measuring and predicting shared situation awareness in teams," *Journal of Cognitive Engineering and Decision Making*, vol. 3, no. 3, pp. 280–308, 2009.

[32] J. Y. Chen, K. Procci, M. Boyce, J. Wright, A. Garcia, and M. Barnes, "Situation awareness based agent transparency," *US Army Research Laboratory*, no. April, pp. 1–29, 2014.

[33] J. Y. Chen, S. G. Lakhmani, K. Stowers, A. R. Selkowitz, J. L. Wright, and M. Barnes, "Situation awareness-based agent transparency and human-autonomy teaming effectiveness," *Theoretical Issues in Ergonomics Science*, vol. 19, no. 3, pp. 259–282, 2018.

[34] L. J. Sorensen and N. A. Stanton, "Keeping it together: The role of transactional situation awareness in team performance," *International Journal of Industrial Ergonomics*, vol. 53, pp. 267–273, 2016.

[35] M. R. Endsley, "Supporting human-ai teams: Transparency, explainability, and situation awareness," *Computers in Human Behavior*, vol. 140, p. 107574, 2023.

[36] J. Burrell, "How the machine 'thinks': Understanding opacity in machine learning algorithms," *Big Data & Society*, vol. 3, no. 1, p. 2053951715622512, 2016.

[37] G. Bansal, B. Nushi, E. Kamar, E. Horvitz, and D. S. Weld, "Is the most accurate ai the best teammate? optimizing ai for teamwork," in *Proceedings of the AAAI Conference on Artificial Intelligence*, vol. 35, pp. 11405–11414, 2021.

[38] M. R. Endsley, C. A. Bolstad, D. G. Jones, and J. M. Riley, "Situation awareness oriented design: from user's cognitive requirements to creating effective supporting technologies," in *Proceedings of the Human Factors and Ergonomics Society Annual Meeting*, vol. 47, pp. 268– 272, SAGE Publications Sage CA: Los Angeles, CA, 2003.

[39] C. D. Wickens, B. A. Clegg, A. Z. Vieane, and A. L. Sebok, "Complacency and automation bias in the use of imperfect automation," *Human Factors*, vol. 57, no. 5, pp. 728–739, 2015.

[40] J. D. Lee and K. A. See, "Trust in automation: Designing for appropriate reliance," *Human Factors*, vol. 46, no. 1, pp. 50–80, 2004.

[41] K. A. Hoff and M. Bashir, "Trust in automation: Integrating empirical evidence on factors that influence trust," *Human F*

[42] *actors*, vol. 57, no. 3, pp. 407–434, 2015.

[43] C. J. Johnson, M. Demir, N. J. McNeese, J. C. Gorman, A. T. Wolff, and N. J. Cooke, "The impact of training on human–autonomy team communications and trust calibration," *Human Factors*, vol. 65, no. 7, pp. 1554–1570, 2023.

[44] S. J. Bickley and B. Torgler, "Cognitive architectures for artificial intelligence ethics," *Ai & Society*, vol. 38, no. 2, pp. 501–519, 2023.

[45] R. J. M. Boyles, "Can't bottom-up artificial moral agents make moral judgements?," *Filosofija. Sociologija*, vol. 35, no. 1, 2024.

[46] G. Chandler, "Unity of command or unity of effort? rethinking a fundamental principle of war," *Modern War Institute at West Point*, 2023.

[47] Headquarters, Department of the Army, *Army Field Manual 6-0: Commander and Staff Organization and Operations*, 2022.

[48] J. Van Diggelen, M. Neerincx, M. Peeters, and J. M. Schraagen, "Developing effective and resilient human-agent teamwork using team design patterns," *IEEE Intelligent sSystems*, vol. 34, no. 2, pp. 15–24, 2018.

[49] J. van Diggelen and M. Johnson, "Team design patterns," in *Proceedings of the 7th International Conference on Human-Agent Interaction*, pp. 118–126, 2019.

[50] J. Cha, M. Barnes, and J. Y. Chen, "Visualization techniques for transparent human-agent interface designs," tech. rep., Technical Report ARL-TR-8674). US Combat Capabilities Development Command Army Research Laboratory Aberdeen Proving Ground United States, 2019.

[51] J. Hansberger, R. Wood, T. Conner, J. Hansen, J. Nix, and M. Torres, "Designing an interactive mixed reality cockpit for enhanced soldier-vehicle interaction," in *Proceedings of the Ground Vehicle Systems Engineering and Technology Symposium*, pp. 16–18, 2023.

[52] D. Amoroso, "Jus in bello and jus ad bellum arguments against autonomy in weapons systems: A re-appraisal," *Questions of International Law*, vol. 4, p. 43, 2017.

[53] L. A. Letendre, "Lethal autonomous weapon systems: Translating geek speak for lawyers," *International Law Studies*, vol. 96, no. 1, p. 11, 2020.

[54] J. D. Weisz, M. Muller, S. Houde, J. Richards, S. I. Ross, F. Martinez, M. Agarwal, and K. Talamadupula, "Perfection not required? Human-AI partnerships in code translation," in *26th International Conference on Intelligent User Interfaces*, pp. 402–412, 2021.

[55] M. Kay, T. Kola, J. R. Hullman, and S. A. Munson, "When (ish) is my bus? User-centered visualizations of uncertainty in everyday, mobile predictive systems," in *Proceedings of the 2016 CHI Conference on Human Factors in Computing Systems*, pp. 5092–5103, 2016.

[56] S. A. Kusumastuti, K. A. Pollard, A. H. Oiknine, B. Dalangin, T. R. Raber, and B. T. Files, "Practice improves performance of a 2D uncertainty integration task within and across visualizations," *IEEE Transactions on Visualization and Computer Graphics*, 2022.

[57] R. Kosara and J. Mackinlay, "Storytelling: The next step for visualization," *Computer*, vol. 46, no. 5, pp. 44–50, 2013.

[58] E. Segel and J. Heer, "Narrative visualization: Telling stories with data," *IEEE Transactions on Visualization and Computer Graphics*, vol. 16, no. 6, pp. 1139–1148, 2010.

[59] G. Rizzolatti and C. Sinigaglia, "The functional role of the parieto-frontal mirror circuit: interpretations and misinterpretations," *Nature Reviews Neuroscience*, vol. 11, no. 4, pp. 264– 274, 2010.

[60] F. H. Previc, "The neuropsychology of 3-d space.," *Psychological Bulletin*, vol. 124, no. 2, p. 123, 1998.

[61] P. A. Longley, M. F. Goodchild, D. J. Maguire, and D. W. Rhind, *Geographic Information Science and Systems*. John Wiley & Sons, 2015.

[62] M. R. Endsley, "A taxonomy of situation awareness errors," *Human factors in Aviation Operations*, vol. 3, no. 2, pp. 287–292, 1995.

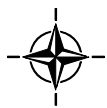